\documentclass[aps,prd,twocolumn,superscriptaddress]{revtex4}
\usepackage{graphicx}
\usepackage{amssymb}

\begin{document}


\title{Results from the First Science Run of the ZEPLIN-III Dark Matter Search Experiment}


\author{V.~N.~Lebedenko\footnote{Deceased}}
\affiliation{Blackett
Laboratory, Imperial College London, UK}
\author{H.~M.~Ara\'ujo}
\affiliation{Blackett Laboratory, Imperial College London, UK}
\affiliation{Particle Physics Department, Rutherford Appleton
Laboratory, Chilton, UK}
\author{E.~J.~Barnes}
\affiliation {School of Physics and Astronomy, SUPA, University of
Edinburgh, UK}
\author{A.~Bewick}
\affiliation{Blackett Laboratory,
Imperial College London, UK}
\author{R.~Cashmore}
\affiliation{Brasenose College, University of Oxford, UK}
\author{V.~Chepel}
\affiliation{LIP--Coimbra \& Department of Physics of the University
of Coimbra, Portugal}
\author{A.~Currie}
\affiliation{Blackett Laboratory, Imperial College London, UK}
\author{D.~Davidge}
\affiliation{Blackett Laboratory, Imperial College London, UK}
\author{J.~Dawson}
\affiliation{Blackett Laboratory, Imperial College London, UK}
\author{T.~Durkin}
\affiliation{Particle Physics Department, Rutherford Appleton
Laboratory, Chilton, UK}
\author{B.~Edwards}
\affiliation{Blackett Laboratory, Imperial College London, UK}
\affiliation{Particle Physics Department, Rutherford Appleton
Laboratory, Chilton, UK}
\author{C.~Ghag}
\affiliation {School of Physics and Astronomy, SUPA, University of
Edinburgh, UK}
\author{M.~Horn}
\affiliation{Blackett Laboratory, Imperial College London, UK}
\author{A.~S.~Howard}
\affiliation{Blackett Laboratory, Imperial College London, UK}
\author{A.~J.~Hughes}
\affiliation{Particle Physics Department, Rutherford Appleton
Laboratory, Chilton, UK}
\author{W.~G.~Jones}
\affiliation{Blackett Laboratory, Imperial College London, UK}
\author{M.~Joshi}
\affiliation{Blackett Laboratory, Imperial College London, UK}
\author{G.~E.~Kalmus}
\affiliation{Particle Physics Department, Rutherford Appleton
Laboratory, Chilton, UK}
\author{A.~G.~Kovalenko}
\affiliation{Institute for Theoretical and Experimental Physics,
Moscow, Russia}
\author{A.~Lindote}
\affiliation{LIP--Coimbra \& Department of Physics of the University
of Coimbra, Portugal}
\author{I.~Liubarsky}
\affiliation{Blackett Laboratory, Imperial College London, UK}
\author{M.~I.~Lopes}
\affiliation{LIP--Coimbra \& Department of Physics of the University
of Coimbra, Portugal}
\author{R.~L\"{u}scher}
\affiliation{Particle Physics Department, Rutherford Appleton
Laboratory, Chilton, UK}
\author{P.~Majewski}
\affiliation{Particle Physics Department, Rutherford Appleton
Laboratory, Chilton, UK}
\author{A.~StJ.~Murphy}
\affiliation {School of Physics and Astronomy, SUPA, University of
Edinburgh, UK}
\author{F.~Neves}
\affiliation{LIP--Coimbra \& Department of Physics of the University
of Coimbra, Portugal} \affiliation{Blackett Laboratory, Imperial
College London, UK}
\author{J.~Pinto da Cunha}
\affiliation{LIP--Coimbra \& Department of Physics of the University
of Coimbra, Portugal}
\author{R.~Preece}
\affiliation{Particle Physics Department, Rutherford Appleton
Laboratory, Chilton, UK}
\author{J.~J.~Quenby}
\affiliation{Blackett Laboratory, Imperial College London, UK}
\author{P.~R.~Scovell}
\affiliation {School of Physics and Astronomy, SUPA, University of
Edinburgh, UK}
\author{C.~Silva}
\affiliation{LIP--Coimbra \& Department of Physics of the University
of Coimbra, Portugal}
\author{V.~N.~Solovov}
\affiliation{LIP--Coimbra \& Department of Physics of the University
of Coimbra, Portugal}
\author{N.~J.~T.~Smith}
\affiliation{Particle Physics Department, Rutherford Appleton
Laboratory, Chilton, UK}
\author{P.~F.~Smith}
\affiliation{Particle Physics Department, Rutherford Appleton
Laboratory, Chilton, UK}
\author{V.~N.~Stekhanov}
\affiliation{Institute for Theoretical and Experimental Physics,
Moscow, Russia}
\author{T.~J.~Sumner\footnote{Corresponding author; address: High Energy Physics
Group, Blackett Laboratory, Imperial College London, SW7 2BW, UK.
Email: t.sumner@imperial.ac.uk}}
\affiliation{Blackett Laboratory,
Imperial College London, UK}
\author{C.~Thorne}
\affiliation{Blackett Laboratory, Imperial College London, UK}
\author{R.~J.~Walker}
\affiliation{Blackett Laboratory, Imperial College London, UK}


\date{\today}

\begin{abstract}
The ZEPLIN-III experiment in the Palmer Underground Laboratory at
Boulby uses a 12\,kg two-phase xenon time projection chamber to
search for the weakly interacting massive particles (WIMPs) that may
account for the dark matter of our Galaxy. The detector measures
both scintillation and ionisation produced by radiation interacting
in the liquid to differentiate between the nuclear recoils expected
from WIMPs and the electron recoil background signals down to
$\sim$10~keV nuclear recoil energy.  An analysis of
847\,kg$\cdot$days of data acquired between February $27^{th}$ 2008
and May $20^{th}$ 2008 has excluded a WIMP-nucleon elastic
scattering spin-independent cross-section above $8.1\times
10^{-8}\,$pb at 60\,GeVc$^{-2}$ with a 90\% confidence limit.
 It has also demonstrated that the two-phase xenon technique is capable of better discrimination between electron and nuclear recoils at low-energy than previously achieved by other xenon-based experiments. \end{abstract}


\maketitle

\section{Introduction}
\subsection{Motivation}
Searches for weakly interacting massive particles (WIMPs) are
motivated by the coming together of unification schemes, such as
supersymmetry, which predict new particle species, and extensive
observational evidence which demonstrates the need for additional
non-baryonic gravitational mass within the Universe.  That the WIMPs
of supersymmetry naturally fulfill this need is remarkably
persuasive.  Indeed, WIMPs occur in other frameworks too. As a
generic class of particle they are assumed to only interact
non-gravitationally with baryonic matter via the weak interaction.
Whilst this offers a mechanism for energy transfer and hence
detection, it also implies rather low event rates and energy
deposits: $<$0.1~events/day/kg and $<$50~keV respectively. This
dictates the use of sensitive underground experiments capable of
specifically identifying energy deposits due to elastic scattering
of incoming particles from target nuclei.  ZEPLIN-III is the latest
in a progressive series of instruments designed to push steadily the
sensitivity limits by exploring alternative approaches using
xenon-based targets~\cite{alner05,alner07}.

\subsection{\label{1B}ZEPLIN-III}
ZEPLIN-III is a two-phase (liquid/gas) xenon time-projection chamber
specifically designed to search for dark matter WIMPs. Its design
and performance details have already been presented
elsewhere~\cite{araujo06, akimov07} and only a brief reminder is
given here. The experiment is operating 1100~m underground.  The
active volume is a disc of 35~mm thickness and $\sim$190~mm diameter which contains $\sim$12~kg of liquid xenon  above an array of
31 2-inch diameter photomultipliers (PMTs). The PMTs employed during
this first science run were ETL D730/9829Q~\cite{araujo04}, and they
were used to record both the rapid scintillation signal, S1, and a
delayed second signal, S2, produced by proportional
electroluminescence in the gas phase above the liquid~\cite{dolg70}.
The PMT array was immersed in the liquid viewing upwards.  The
electric field in the target volume was defined by a cathode wire
grid 36~mm below the liquid surface and an anode plate 4~mm above
the surface in the gas phase. These two electrodes alone produce the drift field in the liquid (3.9~kV/cm), the field for extraction of the charge from the surface, and the electroluminescence field in the gas (7.8~kV/cm). A fiducial volume
for WIMP searches was defined by using a time window for delays
between S1 and S2, which selected a depth slice within the liquid,
and by 2-D position reconstruction from the PMT signals to select a
radial boundary at 150~mm. The time window was set between 500~ns and 13,000~ns which selected depths between 1.29~mm and 33.43~mm.  These together defined a fiducial volume containing 6.5~kg of xenon.
\\
The PMT signals were digitised at 2\,ns sampling over a time segment
of $\pm18\,\mu$s either side of the trigger point. Each PMT signal
was fed into two 8-bit digitisers (ACQIRIS DC265) with a $\times10$
gain difference between them provided by fast amplifiers (Phillips
Scientific 770), to obtain both high and low sensitivity read-out
covering a wide dynamic range.  The PMT array was operated from a common
HV supply with attenuators (Phillips Scientific 804) used to
normalise their individual gains.  The trigger was created from the
shaped sum signal of all the PMTs.  For nuclear recoil interactions
the trigger was always caused by an S2 signal for energies up to
S1=40~keVee, where keVee is an energy unit referenced to the equivalent
S1 signal produced by 122~keV $\gamma$-rays from $^{57}$Co. The
trigger threshold was $\sim$11 ionisation electrons and this
corresponded to $\sim$0.2~keV for electron recoils (for nuclear
recoils see Section~\ref{ec}).  This S2 threshold was set to avoid
excessive triggers from single electron emission events and from
electron and nuclear recoils whose primaries would otherwise have
been undetectable as they fall below the S1 detection threshold.
\\
The xenon target was contained within a vessel itself located within
a vacuum jacket both made from low-background oxygen-free copper.
Cooling was provided by a $40\,$litre liquid nitrogen reservoir,
also made from copper, inside the vacuum jacket. Thermal stability
to $<$0.5~$\rm{^o}$C was achieved over the entire run by controlling
the flow of cold nitrogen boil-off gas through the base-flange of
the xenon vessel. Pressure stability to 2\% was maintained.  The
ZEPLIN-III detector was completely surrounded by a shield of
$30\,$cm thick polypropylene and $20\,$cm thick lead, giving $10^5$
attenuation factors for both $\gamma$-rays and neutrons from the
cavern walls.  Dedicated access through the shield was provided for
the radioactive calibration source delivery, instrument levelling
screws and pipe-work to the external gas purification system.

\subsection{\label{1C}Science Data}

\begin{figure}
\includegraphics[width=3.5in,clip=]{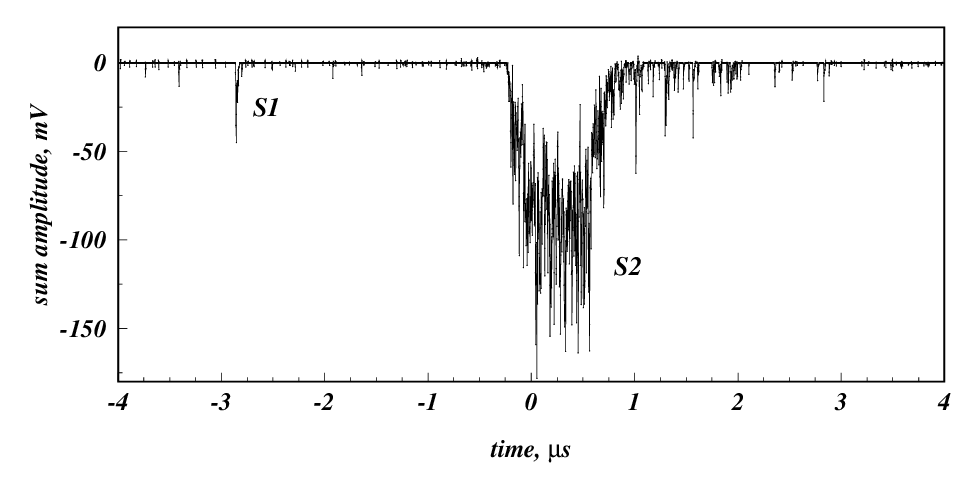}
\caption{\label{event}Segment of the high-sensitivity summed
waveform for a neutron elastic scattering event with S1$= 5~$keVee, showing a small
primary pulse (S1) preceding a large secondary pulse (S2).  Some PMT
after-pulsing and, possibly, single electron emission can be seen
following S2.  Note that only excursions $>$3~{\it rms} on
individual channels are added into the summed waveform.  See later
text for more detailed discussion of some of these points.}
\end{figure}

WIMP-search data were collected over 83~days of continuous operation
in the Boulby Laboratory starting on $27^{th}$ February 2008. An
84\% live time was achieved during the science run and some
$847\,$kg$\cdot$days of raw data were collected from the $12\,$kg
target volume.  $^{57}$Co calibration measurements were made every
day. Nuclear recoil calibrations were made with an AmBe neutron
source at the beginning and end of the 83~day period (5~hrs each). A
typical event, from a neutron elastic scattering interaction in the liquid
with S1=5~keVee, is shown in Figure~\ref{event} as
recorded through the high-sensitivity sum channel.  A short Compton
calibration was performed using a $^{137}$Cs source at the beginning
of the run with a much longer run at the end (122~hrs). Ten percent of the science data (every $10^{th}$ file) were used to develop initial data analysis and selection cuts, to establish the level of the electron-recoil background, and to define the boundaries for the WIMP-search box and its acceptance. At first, the remaining 90\% of the science data were retained unopened to carry out a `blind' analysis, but these data were eventually used for perfecting some data-selection cuts as detailed below, making the final analysis non-blind.\\
\\
Pulse-finding algorithms were used to identify signals in the 62
waveforms (independently for each PMT and for high and low
sensitivity channels).  These were then categorised as S1 or S2
candidates based on a pulse width parameter (charge mean arrival
time, $\tau$): scintillation pulses are much shorter
($\tau$$\lesssim$40~ns) than electroluminescence pulses, with
durations corresponding to the drift time across the gas gap
($\tau$$\sim$550~ns).  Viable S1 and S2 candidates were then subject
to software thresholds ($\ge$3 channels recording signals above
$1/3$~photoelectron (p.e.) for S1 and a minimum area of
$\sim$5~ionisation electrons for S2).  Only events with one S1 and
one S2  were considered for further analysis.
Of particular note here, $\chi^2$ goodness of fit indicators within
the position reconstruction of both S1 and S2 were used to remove
multiple-scatter events, and this was particularly effective for
those with one vertex in a `dead' region of the xenon, which would
otherwise have been a troublesome background.  Such `dead' regions
include the reverse-field volume between the cathode wire and the
PMT grid wire \cite{akimov07} and the thin (0.5~mm) layer of xenon
surrounding the PMT bodies.  Double-Compton interactions with at
least one vertex in these regions, referred to as
`multiple-scintillation single-ionisation' (MSSI) events, fulfil the
previous selection criteria since there is no S2 pulse from the dead
region and the coincident scintillation pulses are added together in
a single S1. Unfortunately, perfecting this selection eventually
required use of the full data-set as will be described in more detail
below. \section{Calibration}
\subsection{Scintillation Response and Position Reconstruction\label{co57_cal}}
\begin{figure}
\includegraphics[width=3.5in,clip=]{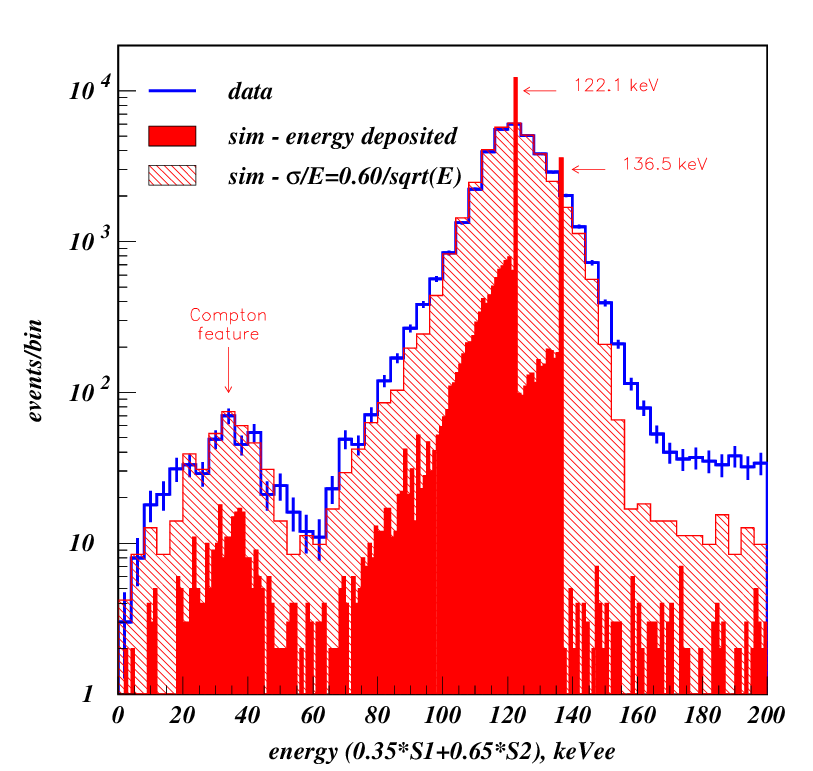}
\caption{\label{co57}Response to an external $^{57}$Co $\gamma$-ray
source in the combined energy channel, exploiting S1 and S2
anti-correlation. One day's experimental data are shown in blue with
statistical error bars. The simulation result is indicated in red:
the solid histogram shows the bare energy deposits and the shaded
one shows the result of Gaussian-smearing with the energy resolution
indicated in the figure.}
\end{figure}

\begin{figure}
\includegraphics[width=1.65in,clip=]{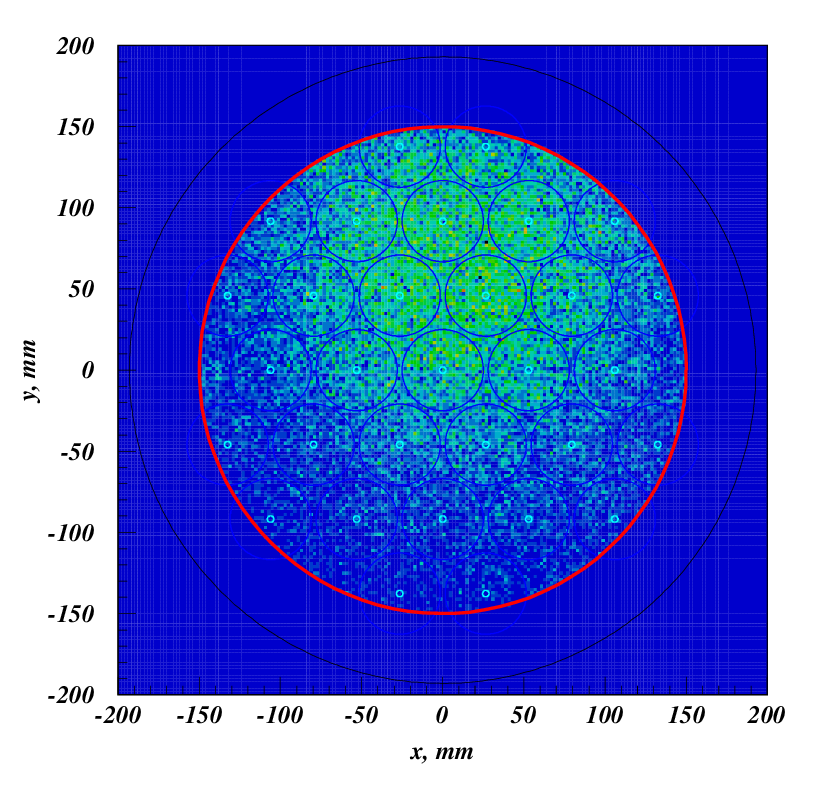}
\includegraphics[width=1.65in,clip=]{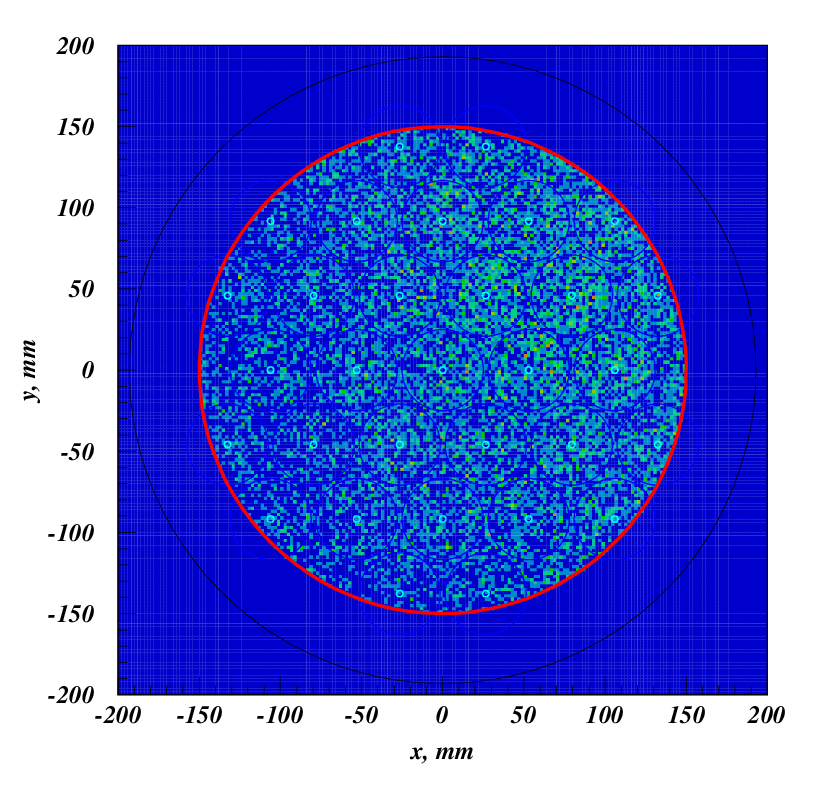}
\caption{\label{co57_pos}Distribution in the horizontal plane of
events from the $^{57}$Co source on the left and from the AmBe source on the right. The source positions are different for each image and neither is centred.   In both cases the volume distribution is as expected from Monte Carlo simulations, given the location of each source. Interaction vertices can be seen out to the edge of
the fiducial volume at a radius of 150~mm (red circle). The outer
circle shows the edge of the liquid xenon target. Each PMT is marked
by two smaller circles (PMT centres and envelopes).}
\end{figure}

An external $^{57}$Co source was inserted through the shield and
located above the instrument every day. The dominant 122~keV
$\gamma$-rays have a photoelectric absorption length of 3.3~mm in
liquid Xe, and hence provided good standard calibration candles from
interactions close to the liquid surface.  A typical $^{57}$Co
spectrum is shown in Figure~\ref{co57}.  The S1 signal channel
exhibited a light detection efficiency at our operating field
(3.9~kV/cm) of $L_y$=1.8~p.e./keVee, decreasing from 5.0~p.e./keVee on
application of the electric field.  The 122~keV interactions were
used for a number of purposes to calibrate the instrument.
Using S2 pulses, an iterative procedure, whereby a common
cylindrical response profile was fitted to each channel, was used to
normalise the measured response from each PMT (i.e.~`flat-field' the
array). Position reconstruction in the horizontal plane was then
achieved by using the converged response profiles in a simultaneous
least-squares minimisation to all channels~\cite{solo08}.  This
method complements the Monte Carlo template matching procedure also
being used but is less dependent on accurate iterative
simulations~\cite{lindote07}.
Finally, the integrated areas of the S1 and S2 responses gave light
collection correction factors as a function of radial position.
Using this procedure a full-volume energy resolution of
$\sigma$=5.4\% at 122~keV was obtained with an energy reconstruction
using a combination of the S1 and S2 responses to reflect the fact
that, for electron recoils, these two channels are anti-correlated
at a microscopic level.  The individual S1 and S2 resolutions at
122~keV are 16.3\% and 8.8\%, respectively.  Also shown in
Figure~\ref{co57} is the comparison of the response to simulation.
Not only are the two main $^{57}$Co lines well fitted but there is
also a good match to the predicted Compton feature at $\sim$35~keV.
The excess above 150~keV is mainly due to the unsubtracted
background. The left-hand panel in Figure~\ref{co57_pos} shows the distribution in the $x$-$y$ plane of
events seen from the source. As expected most events are located
towards the centre (the offset is due to an offset source position)
with a radial fall-off as expected from the increasing thickness of copper
along the line of sight.
\subsection{Stability, Electron Lifetime and Detector Tilt}
The $^{57}$Co daily calibrations were used to assess the evolution
of other operational parameters over the entire run: i) the average
light and ionisation yields, as measured by fits to the $^{57}$Co S1
and S2 pulse area spectra; ii) the mean electron lifetime in the
liquid, obtained from the exponential depth dependence of the ratio
of the areas of the S2 and S1 signals (hereafter simply referred to
as $S2/S1$); iii) the evolution of the long-term detector tilt due
to local geological factors, as given by the polar dependence of the
S2-width distribution, which probes the thickness of the gas layer.
The detector tilted by less than 1~mrad over the run.  Over the fiducial volume this corresponds to a systematic change in the gas gap of $<3.5$\%,  which in turn translates proportionally into a variation in the S2 signal. This was not
deemed sufficient to warrant a full correction~\footnote{This has been verified by subsequent direct analysis and details will be published elsewhere}. The scintillation
mean light yield remained stable to a few percent, as did the
ionisation yield, after correcting for the electron lifetime in the
liquid.  Remarkably, the lifetime did show an evolution during the
run in the form of an improvement: from an initial value of
20~$\mu$s, achieved by initial gas-phase purification through
external getters, a value of 35~$\mu$s had been reached by the end
of the run (the full drift length of the chamber is only 14$\,\mu$s).
There was no active recirculation used and this improvement is
attributed to the clean, xenon-friendly materials used in detector
construction and to the uninterrupted application of the electric
fields during the entire run. As the area ratio $S2/S1$ is the main
discriminant between nuclear and electron recoils, a depth-dependent
correction must be applied to the S2 area to compensate for electron
trapping by impurities. The electrons from the deepest events within the fiducial volume drifted for 13$\,\mu$s and the correction factor for these varied from 1.92 at the start of the run to 1.45 at the end. The daily $^{57}$Co calibrations allowed
this to be monitored throughout the science run and events were
corrected individually using an historical trend profile.

\subsection{Linearity}
The linearity of the response of each channel in the array was
investigated using low-energy Compton-scattered events from the
$^{137}$Cs source, in order to rule out hardware and software
distortion for processing of small signals. The position of the vertex for each interaction was found and the waveforms from PMTs located a
certain distance away from the vertex were selected
based on the expected number of S1 photons, given the cylindrical response profile determined from the $^{57}$Co data as pointed out in \ref{co57_cal}. Provided that the expected
number is indeed small, the mean of the Poisson distribution for the
number of detected photons can be quite accurately determined by
counting the fraction of waveforms which do not contain any identified pulses, i.e. the frequency characterising
the absence of any signal. This assertion is made against a sample
of pure noise in the same waveform. Repeating this procedure for all
channels and a range of expected signal allowed comparison of the
mean S1 pulse area recorded in each trial against the expected
Poisson mean, as shown in Figure~\ref{pmtlin} for the central PMT.
In addition, this provides a very robust method to obtain the mean
size of one photoelectron \cite{nev08}.
This has been calculated for every PMT within the array:
\begin{figure}
\includegraphics[width=3.5in,clip=]{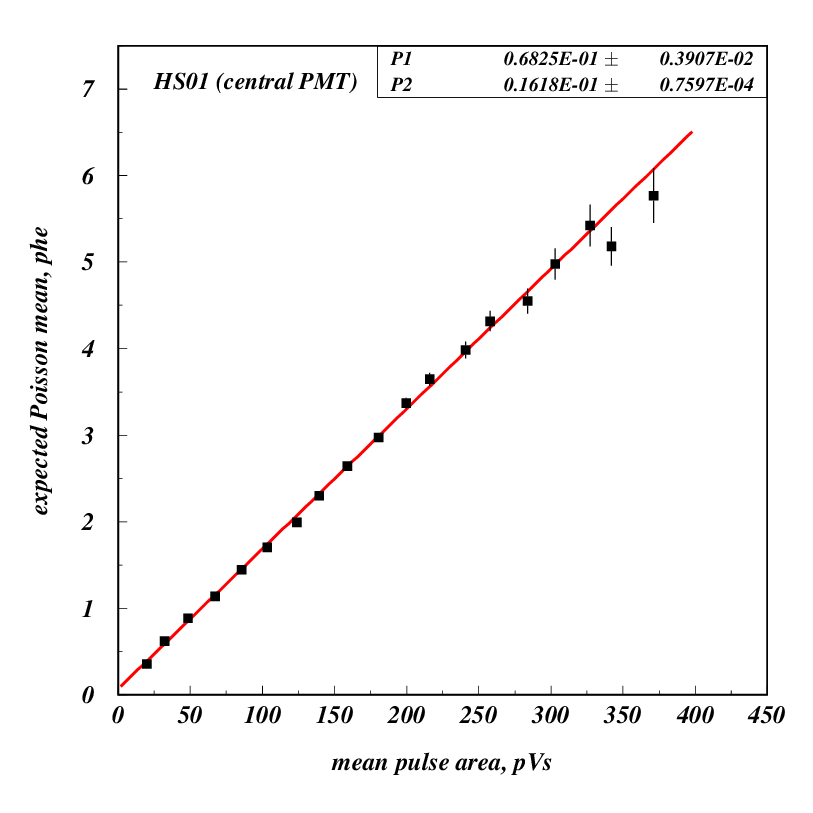}
\caption{\label{pmtlin}Expected mean number of S1 photoelectrons as
a function of the mean pulse area observed in the central channel in
the array. The expected signal is the mean of the Poisson
distribution obtained by counting the frequency of `zeros', i.e. the
absence of any response.}
\end{figure}
the relationship is found to be linear to within the statistical
accuracy of the measurement over a factor of 10 in mean pulse area,
which covers the range of interest for WIMP nuclear recoil signals.
The slope of the line in Figure~\ref{pmtlin} provides a measure of
the mean single photoelectron response (SER) for that PMT. The mean SER
of all the PMTs in the array has been found in this way to be in
the range $47\pm12$~pVs. The spread in these values forms part of
the `flat-field' correction discussed earlier; other dominant
factors are the PMT quantum efficiency and imperfect hardware
equalisation.
\subsection{Nuclear Recoil Response}
The nuclear recoil response in the energy range of interest to WIMP
signals has been calibrated with neutrons from an AmBe ($\alpha$,n)
source. The source was placed inside the polypropylene shielding
above the detector but displaced to one side to reduce the
interaction rate. The right-hand panel in Figure~\ref{co57_pos} shows the reconstructed
event positions from the second calibration performed just after the
science run had been completed. The distribution is slightly
non-unform in the $x$-$y$ plane as expected.

\begin{figure}
\includegraphics[width=3.5in,clip=]{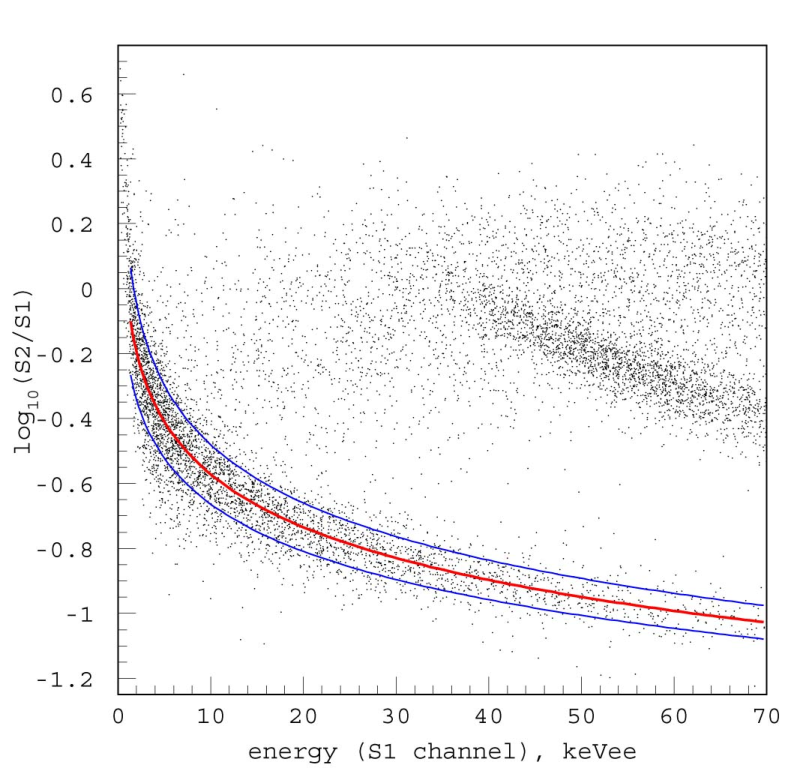}
\caption{\label{ambe_s2s1}Calibration of the nuclear recoil response
with an AmBe neutron source, plotted as the discrimination parameter
($\log_{10}\left(S2/S1\right)$) as a function of
`electron-equivalent energy' (i.e. using the S1 channel calibrated
by $^{57}$Co). The lines show the trends of the mean and standard
deviation of energy-binned log-normal fits to the recoil population.
The distinct population above $\sim$40~keVee is due to inelastic
neutron scattering off $^{129}$Xe nuclei.}
\end{figure}

\begin{figure*}
\includegraphics[width=7in,clip=]{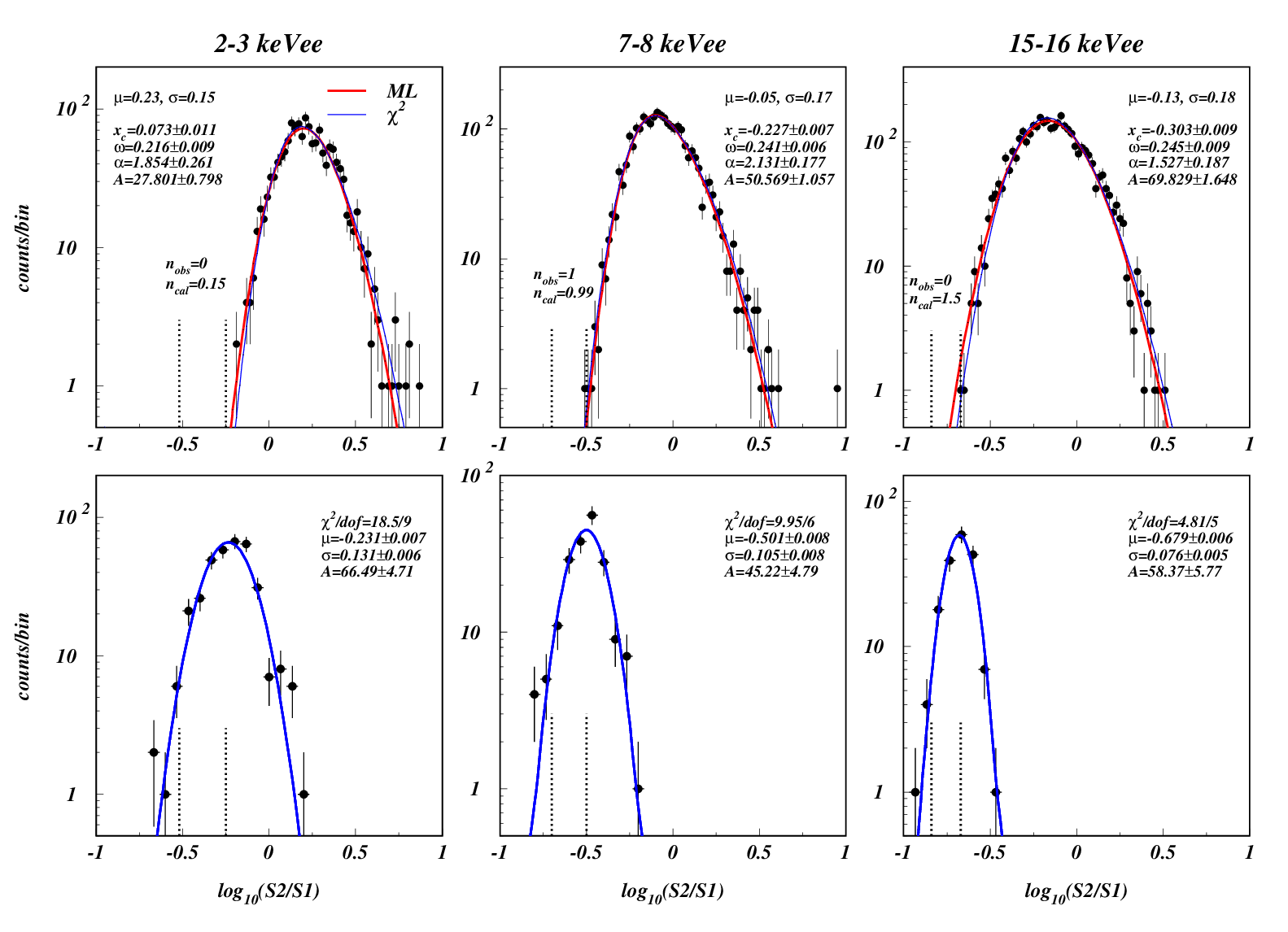}
\caption{\label{ambe_slice}Statistical fitting of the electron and
nuclear recoil populations using the WIMP-search (upper panel) and
AmBe data-sets (lower panel). Three 1-keVee wide bins are shown:
lowest, intermediate and highest energies accepted. The electron
recoil population was fitted with a skew-Gaussian function using
both the minimum $\chi^2$ (thin blue line) and a maximum likelihood
(ML) method with the Poisson distribution as estimator (thick red line). The latter fit is more appropriate to data with low statistics (including zeros)\ in the tails of the populations as it uses a Poisson distribution as an estimator. Note that the entire population can be fitted (all energy bins, across the entire $\log_{10}(S2/S1)$ range).
The ML best fit parameters are indicated, along with the mean and
standard deviation of the skew-Gaussian. The lower panels show the
log-normal fits to the AmBe recoil data, which is used to define the
acceptance region [$\mu$--$2\sigma$,$\mu$], between the vertical
dashed lines. The number of electron recoils
observed to be leaking into this region, $n_{obs}$, is compared with the estimated
number, $n_{cal}$, from the ML fits.  The total number of events
expected in the acceptance region is 11.6$\pm$3.0.}
\end{figure*}

\begin{figure}
\includegraphics[width=3.5in,clip=]{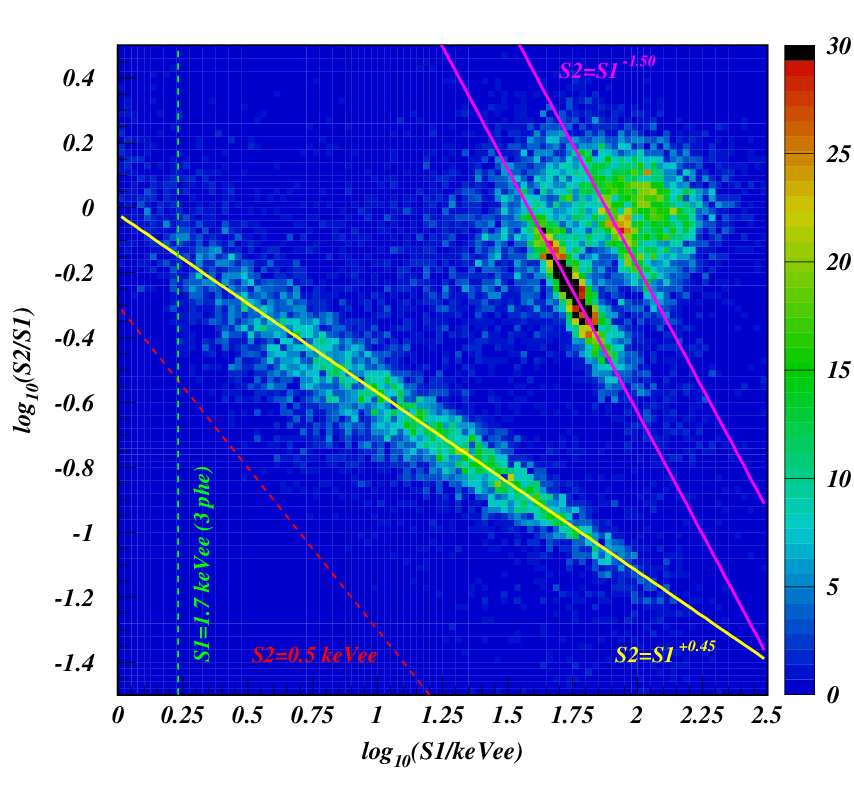}
\caption{\label{ambe_s2s1log}Double-logarithmic plot of
Fig.~\ref{ambe_s2s1} showing the nuclear recoil population obeying
the power-law trend indicated by the yellow line; the behaviour of
the inelastic line from $^{129}$Xe is markedly different, as this is
dominated by charge recombination of the 40~keV $\gamma$-ray rather
than the small nuclear recoil component of the deposited energy.
Approximate thresholds for S1 (3-fold software trigger) and for S2
(hardware trigger) are also indicated.  From this it can be seen that the S2 hardware trigger corresponds to S1=0.5\,keVee for nuclear recoils.}
\end{figure}

Figure~\ref{ambe_s2s1} shows a `scatter-plot' of $\log_{10} (S2/S1)$
as a function of energy in keVee from the AmBe calibration. The red
line shows a smooth fit to the median of the elastic scatter
distribution with $\pm1\sigma$ boundaries as blue lines. To obtain
these curves the data were histogramed into 1~keVee bins and fitted
by log-normal distributions. Examples of the quality of the fits are
shown in Figure~\ref{ambe_slice}. The other well defined population
in Figure~\ref{ambe_s2s1}, between 40--70~keVee, is due to inelastic
scattering of neutrons from $^{129}$Xe nuclei and the more diffuse
horizontal population is caused by associated $\gamma$-ray
interactions. The elastic nuclear recoil median turns out to be very
closely approximated by a power law, which is shown most effectively
by replotting the figure in log-log form
(Figure~\ref{ambe_s2s1log}). Not only is the power-law behaviour
very apparent but it can also be seen that there is less obvious
flaring at lower energies than seen in other xenon experiments whose
data were taken at much lower electric
fields~\cite{alner07,angle08}. Also shown are lines illustrating the
approximate thresholds for S1 and S2.

\subsection{Electron Recoil Response}
\begin{figure}
\includegraphics[width=3.5in,clip=]{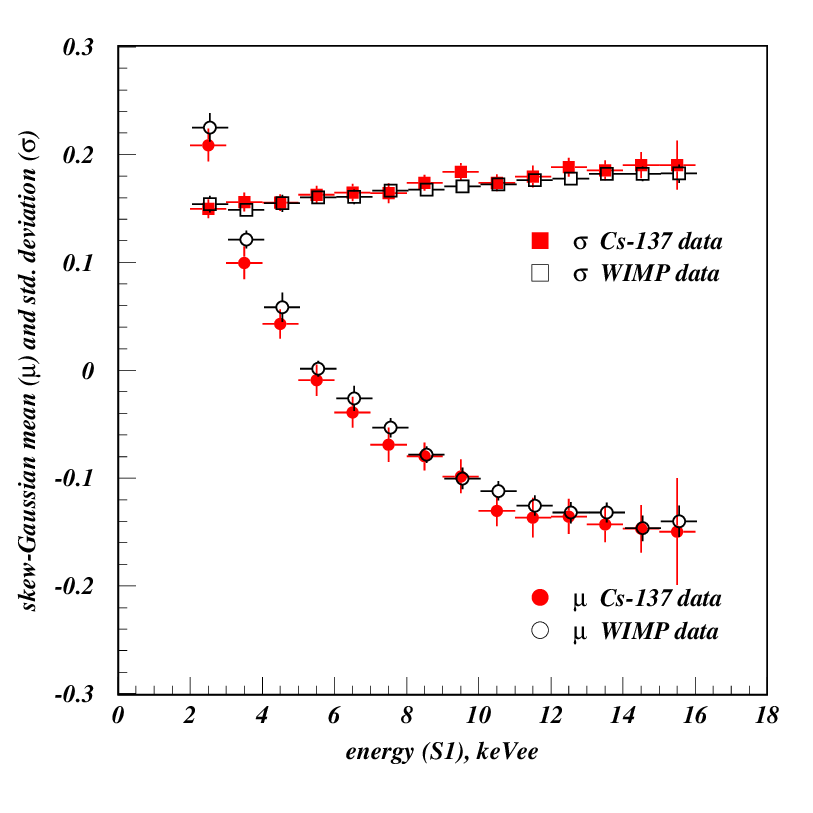}
\caption{\label{cs137_scatter}Comparison of the skew-Gaussian mean
and standard deviations for the $^{137}$Cs and WIMP-search data-sets
calculated from the ML-fit parameters (the horizontal error bars
indicate the bin width).}
\end{figure}

The electron recoil response at low energies was established using a
long duration calibration with a $^{137}$Cs radioactive source.
Compton scattering of the $662$~keV $\gamma$-rays produced a
significant number of events down to $\sim2$~keVee but with only a
small number extending far enough down in the $S2/S1$ parameter to
reach the nuclear recoil median (Figure~\ref{cs137_scatter}). The
general behaviour of the electron recoil band is reminiscent of the
XENON10 results \cite{shutt07,angle08,sore08}, but with a slightly
more pronounced upturn at low energy, a larger separation between
electron and nuclear recoil bands and narrower distributions.
The low-energy electron-recoil populations in the $^{137}$Cs and the
WIMP-search data-sets were fitted in 1~keVee bins by a skew-Gaussian
function.  The fits were performed using a maximum likelihood (ML)
method with a Poisson distribution as estimator for the observed
data. Three of the fits are shown in Figure~\ref{ambe_slice}. The
distribution parameters are consistent bin-by-bin for the
$^{137}$Cs and WIMP data-sets, as confirmed in
Figure~\ref{cs137_scatter}. However, there are two distinct differences in the general behaviour.  Firstly, the mean of the $^{137}$Cs data is systematically lower than that of the WIMP\ data.  It has been shown that this reduction is due to the high count rate used in collecting the Cs data causing the gain of the PMTs to be slightly suppressed due to saturation effects at low-temperature~\cite{nev08}. However, it  was not feasible to lower the rate and still acquire sufficient data in a reasonable time and uncontaminated by other background.     Secondly, the behaviour of the $^{137}$Cs
data-set in the low $S2/S1$ tails is not closely representative of
the science data, with the former exhibiting significantly more
outliers. These events are attributed to MSSI double-Compton events as had been anticipated in~\cite{araujo06}.  This is not evident in Figure~\ref{cs137_scatter} as the number of events concerned was not sufficient to affect the standard deviation noticeably.   \\

Double-Compton events in which both vertices are within the active
volume produce two primary signals which are time coincident, but
separated in position, and two secondary signals which are separated
in both time (delay) and position.  Even if they cannot be
separated they are of no consequence as the combined ratio of
$S2/S1$ will be relatively unaffected. However, if one of the
vertices occurs in a position from which no secondary is possible,
then the only way to identify them is through positional mismatch
between S1 and S2 and a less well reconstructed position from S1 as
this has two vertices.   If the `dead' vertex is very close to one
of the PMT surfaces the S1 signal can also appear to be too peaked
within the array.  Although there were already specific software
cuts designed to deal with these events, some with certain
topologies were not being fully identified by our analysis at that
stage. For the $^{137}$Cs data this problem was most apparent in the
region $\log_{10}$$(S2/S1)<-0.5$ and $E>30$~keVee but extended
right down to the lowest energies.
The $^{137}$Cs calibration data were thus used to improve our
algorithms for identifying MSSIs and the new routines were
implemented after the science data had been opened.  However, even with the
improved selection cuts it was still not possible to use the
$^{137}$Cs data to predict accurately the expected number of
single-scatter events leaking into the nuclear recoil region.  The combination of lowering of the band mean (due to the rate dependent PMT sensitivity suppression at low temperature) towards the nuclear recoil band and remaining additional events in the lower wing caused a large over-prediction of event leakage into the WIMP search box (41 events were predicted).   The additional events remaining in the lower wing were probably due to the  $^{137}$Cs source not accurately
mimicking that of the background sources due to its location. Hence, instead of using the $^{137}$Cs data, the WIMP-search
data themselves were used to estimate the expected electron-recoil
backgrounds, and this gave 11.6$\pm$3.0.
  
\section{Analysis of the WIMP Search Data}

\subsection{Data processing and selection}
The raw data were reduced using the purpose-developed code ZE3RA
(ZEPLIN-III Reduction and Analysis). The DAQ hardware records the 62
waveforms at 500~MS/s (2~ns samples) for 36~$\mu$s periods. ZE3RA
finds candidate pulses in individual waveforms by searching for
3~{\it rms} excursions above the baseline.  Subsequent waveform
processing includes resolving adjacent/overlapping pulses and
grouping of statistically consistent structures (e.g. scintillation
tails). A statistically-motivated timing/shape coincidence analysis
was then used to correlate occurrences on different channels thus
allowing further pulse interpretation (e.g. clustering,
identification of random coincidences, etc.)  The resulting pulses
were ordered by decreasing area in the high-sensitivity (HS) sum
channel and the largest 10 were stored in databases for further
analysis.  By design, ZE3RA does not ascribe physical meaning to
pulses, it rather parameterises them in terms of arrival time,
width, area, amplitude, etc.  An event browser allows visual
scanning of events, channels or individual pulses; a batch-mode
interface allows scripted reduction of large data-sets.



The data structures produced by ZE3RA were analysed by a flexible
code based on hbook~\cite{hboo}.  It processed the original
parameters to assign physical meaning to pulses in events according
to a well defined set of rules (e.g. primary scintillation signals
are fast and must precede wider electroluminescence signals).  Only
events that can represent single scatters in the two-phase target
(`golden' events with one S1 and one S2) were retained. Primary (S1)
pulses were found by applying an acceptance threshold of 1/3 p.e. to
the ZE3RA pulses and also requiring a 3-fold coincidence amongst the
31 PMTs. This software threshold was nominally equivalent to an
energy threshold of 1.7\,keVee.   Exceptions in the S1 selection
were allowed for PMT after-pulses.  These are signal-induced
artifacts generated within the PMTs.  In general they have a
characteristic time delay from the optical signal, but with a wide
distribution and, moreover, it varies between PMTs.  As a result it
is not trivial to identify after-pulsing and avoid them instead
being classified as additional S1 signals, which would result in the
event being wrongly rejected.
Secondary (S2) pulses were required to have at least an integrated area
corresponding to the signal expected from about 5 electrons leaving
the liquid surface. This suppresses optically-induced
single-electron emission~\cite{edwards08} as well as optical
feedback effects from the cathode grid, which are not part of the
direct measure of the ionisation signal generated at the interaction
site.  Many additional parameters are derived for these, such as 3-D
position information, hit-pattern descriptors, interaction energy
and corrections (e.g. array flat-fielding, electron lifetime, liquid
level, light collection, etc). Subsequent analysis (science
exploitation) is based on PAW~\cite{paw} and ROOT~\cite{root}. \\
Trapping MSSI events effectively was a significant challenge,
involving a combination of approaches: use of goodness of fit
indicators in the position reconstruction algorithms, comparison of
coordinates derived independently from S1 and S2, and searching for
abnormal light patterns across the array.

\subsection{The WIMP Search Box}
\label{cuts}
\begin{figure}
\includegraphics[width=3.5in,clip=]{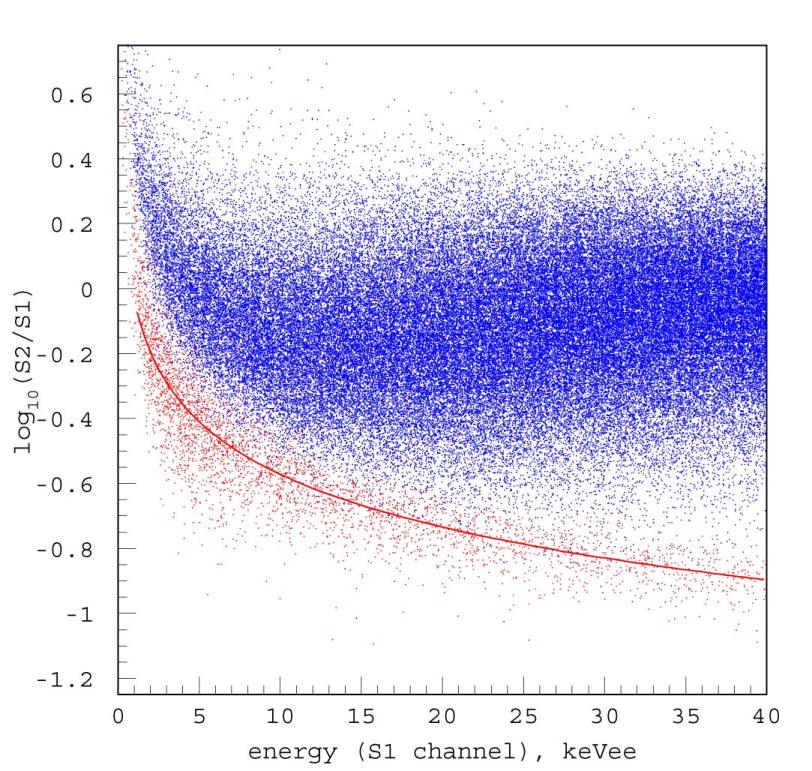}
\caption{\label{disc}Combined scatter plot of $\log_{10}(S2/S1)$ as
a function of energy from the two calibration data-sets, $^{137}$Cs
and AmBe. The upper population corresponds to low-energy Compton
electrons and the narrower, lower one to nuclear recoils produced by
neutron elastic scattering.}
\end{figure}

Discrimination between nuclear and electron recoils is
illustrated in Figure~\ref{disc} which combines electron recoil data
from $^{137}$Cs and elastic nuclear recoil data from AmBe.  The
separation between the two populations is clear and this is used as
the main way of defining the nuclear recoil search box for potential
WIMP events.  The selection cuts used can be categorised as follows:
\begin{enumerate}
\item Golden event selection (including pulse finding, S1 and S2 definition, and single scatter selection)
\item Waveform quality cuts (mild cuts mainly aimed at large baseline excursions compromising pulse parameterisation)
\item Pulse quality cuts (mild cuts to avoid extreme outliers in parameter distributions)
\item Fiducial volume definition (drift time window and a radial limit from the S2 position reconstruction)
\item Event quality cuts (strong cuts to deal with MSSI events mainly)
\end{enumerate}
The fiducial definitions (4) leave an active mass of 6.52\,kg with
a raw exposure of 453.6\,kg.days.
Low-energy events in the 10\% data were well separated from the
nuclear recoil median line down to the lowest energies.  The WIMP
search box boundary was thus defined as $2$$<$$E$$<$16~keVee and
$(\mu_n - 2\sigma)$$<$$\log_{10}(S2/S1)$$<$$\mu_n$, where $\mu_n$ is
the energy-dependent mean of the nuclear recoils (acceptance of
47.7\%). This region was defined before unblinding and was kept for
the subsequent analysis. The effective total exposure within this
box, after taking account of all of the efficiencies, as detailed in
Table~\ref{efftab}, is 127.8~kg$\cdot$days.
\subsection{Backgrounds}
\begin{figure}
\includegraphics[width=3.5in,clip=]{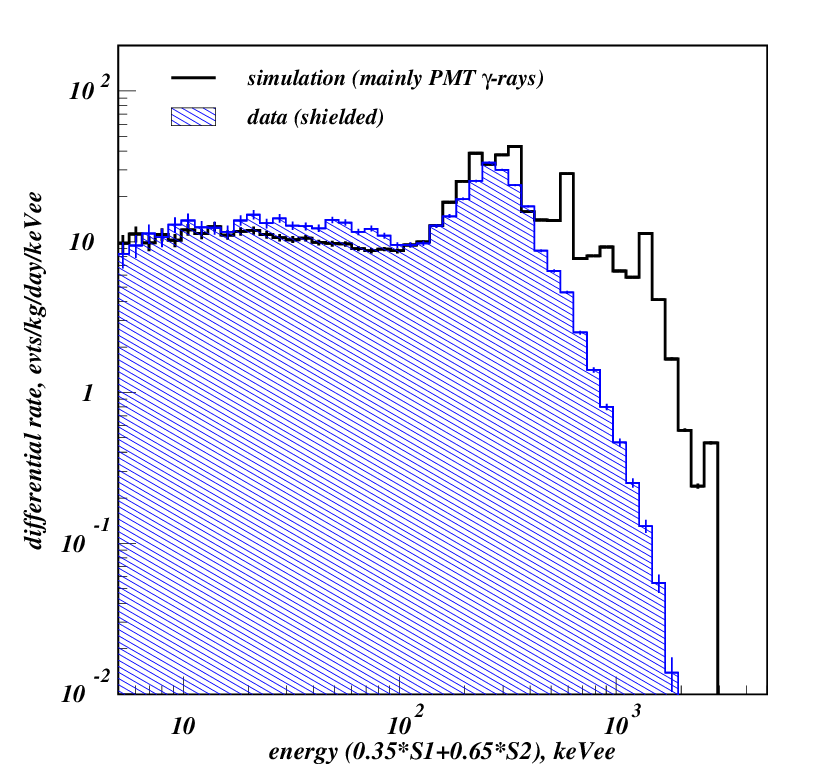}
\caption{\label{gback}Electron recoil background measured during the
fully-shielded science run. The differential spectrum is shown
superimposed on the Monte Carlo prediction~\cite{araujo06} using
GEANT4~\cite{ago03} without rescaling. The latter includes a
dominant 10.5~evts/kg/day/keVee (`dru') from the photomultipliers,
$\gamma$-rays from the lead `castle' (0.7~dru), $\beta$-particles
from $^{85}$Kr (0.2~dru) and $\gamma$-rays from ceramic feedthroughs
(0.1 dru). The disagreement at high energies is caused by
single-scatter selection in the data (but not in the simulation) and
by the limited DAQ dynamic range which was optimised for the
WIMP-search run.}
\end{figure}

Electron and nuclear recoil background predictions for ZEPLIN-III
are based on a full GEANT4~\cite{ago03} simulation including
measured radioactive content levels for all major components\cite{araujo06}.  The
largest contributor, by far, is the PMT array.  Figure~\ref{gback}
shows the measured differential background spectrum together with
the simulated background.  The high-energy region above 300~keVee is
suppressed due to dynamic range limitation.

The expected single-scatter neutron background in the data-set is $1.2\pm0.6$ in the
WIMP search box with 90\% coming from PMT generated events
through ($\alpha$,n) interactions and spontaneous fission.  The remaining 10\%
are mainly from contaminants in ceramic feedthroughs and external leakage through the shield of neutrons from the rock.

\subsection{WIMP Signal Search}
\begin{figure}
\includegraphics[width=3.5in,clip=]{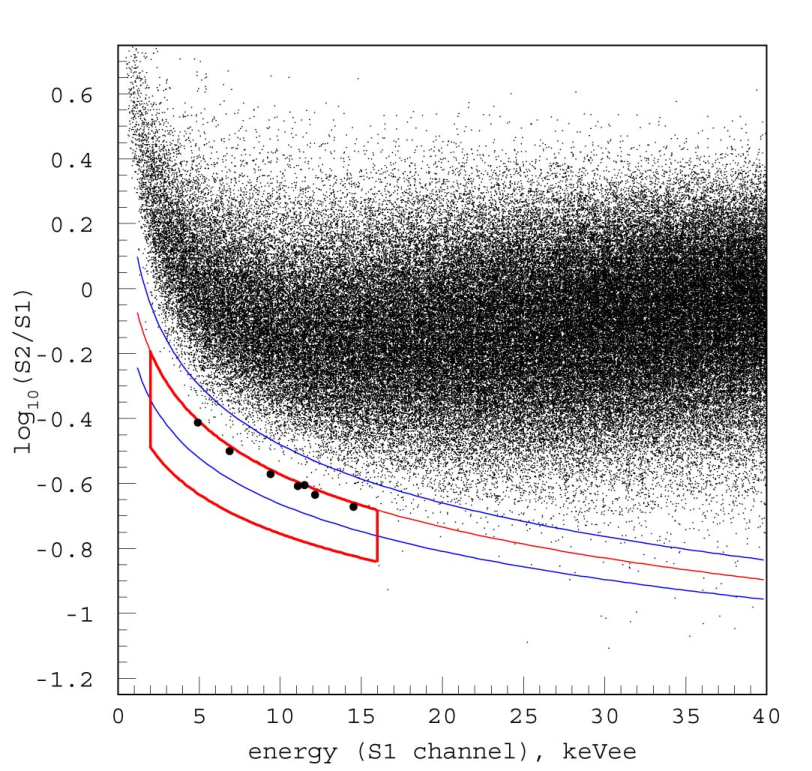}
\caption{\label{fsr}Scatter plot of $\log_{10}(S2/S1)$ as a function
of energy for the entire 83-day data-set of first science run. There
are 7 events (large dots) in the WIMP-search region (thick red box), which extends from $2<E<16$~keVee and $\mu_n-2\sigma_n<\log_{10}(S2/S1)<\mu_n$, where $\mu_n$ is the energy-dependent mean of the nuclear recoils (thin red line bordered by the blue curves at $\pm1\sigma_n$). These are all located near the upper boundary, between
$\simeq$5--15~keVee.}
\end{figure}
Figure~\ref{fsr} shows the final scatter plot from the complete
science data-set.  There are 7 events within the WIMP search box and
the energy scale is shown in keVee.  To assess the implications of
these events the energy scale needs to be converted into keVnr, the
energy dependent detector efficiency for nuclear recoils must be
found and the relative likelihood of any of those 7 events being
drawn from the expected WIMP distribution rather than the extended
electron-recoil distribution must be calculated.

The level of discrimination apparent in Figure~\ref{fsr} is very
high. As derived from the data themselves, the average $\gamma$-ray
rejection factor is 5$\times$10$^3$ between 2--16~keVee with an
increase below 5~keVee.  This is significantly better than had previously been demonstrated by the XENON10 experiment~which achieved 99.9\% at the very lowest energies~\cite{angle08} whilst our data exhibit better than 99.99\% in the 2-5\,keVee band.  

\begin{figure}
\includegraphics[width=3.5in,clip=]{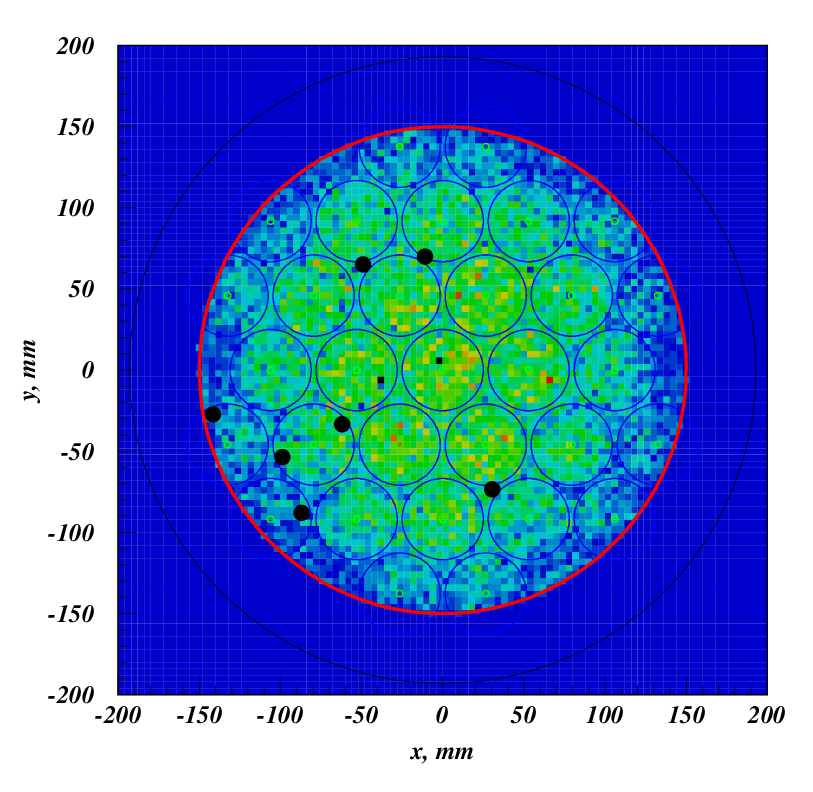}
\caption{\label{rad_wimps}Horizontal distribution of events in the
energy range 2-16~keVee for the science run data-set produced in the same way as those in Figure~\ref{co57_pos}. The
reconstructed location of the 7 events in the acceptance region is
indicated. The measured distribution of the overall background is consistent with detailed Monte Carlo simulations of that expected from the instrument activity which is dominated by the photomultipliers.  }
\end{figure}

Figure~\ref{rad_wimps} shows the spatial $x$-$y$ distribution of all
events in the 2--16~keVee energy range.  Events within the WIMP
search box are highlighted.

To derive the significance of the events within the search box the experiment efficiency must be derived together with the energy scale conversion between keVee and keVnr.  These are established in the following sections.  First the efficiency for nuclear recoil detections is found by comparing AmBe data-sets with very different trigger thresholds in both hardware and software.  The energy scale conversion is then done by comparing a simulation of the expected nuclear recoil with that measured. 
\subsubsection{Efficiency and threshold}
The overall detection efficiency will be a combination of
hardware and software effects.  As mentioned earlier in Section~\ref{1B} the hardware
trigger threshold is derived from S2 in the low-energy part of the
S1 spectrum relevant to WIMP signals.
At higher energies, well beyond the upper limit of the
WIMP search box, there is a high-level inhibit to suppress the
overall count rate, but this does not affect the efficiency at low energy. Dead-time effects are usually energy
independent. Software effects include thresholding associated with pulse finding
algorithms and selection cuts.  These have been described in Section~\ref{1C}.    The energies (expressed in S1~keVee) at which these `thresholds' will affect the detection efficiency are tabulated in Table~\ref{efftab}.  In order to confirm these expectations a second AmBe data-set was analysed as a check on the energy dependence near the threshold.  This data-set had been acquired with a lower
hardware S2 trigger threshold.  In addition, the 3-fold S1 coincidence
requirement was changed to 2-fold in this particular analysis and all quality
cuts removed or significantly relaxed. The overall effect of these two changes is
shown in Figure~\ref{ambe_spec} by comparing the black histogram labelled `Am-Be low-threshold data' with the blue shaded histogram labelled `Am-Be calibration data'. The difference between these two histograms  is only noticeable below S1$\sim$4~keVee as expected.
A study of the smallest S2 events triggering the system
in each run has shown directly that the trigger level in the two runs was
$\sim$11 and $\sim$4 ionisation electrons, respectively.  These
numbers were calibrated against the measured single electron
spectrum for ZEPLIN-III following the method already used for
ZEPLIN-II~\cite{edwards08}. The experiment efficiency during the science run is taken as the ratio of the two AmBe data-sets, shown in Figure~\ref{nlin}.

The full red curve labelled `Simulation (Eee/Enr=2.09)' in Figure~\ref{ambe_spec} shows  a Monte Carlo simulation of the expected differential spectrum which should have been seen by the experiment assuming a constant ratio between S1~keVee and S1~keVnr.   This simulated curve has not been corrected for instrument efficiency but even so it is clear that there is a departure from the experimental data below S1~$\sim20$~keV.  Given that this mismatch extends so far in energy above any reasonable thresholding effects, it is interpreted as evidence for a non-linear scale conversion.
\subsubsection{Energy conversion}
\label{ec} 
\begin{figure}
\includegraphics[width=3.5in,clip=]{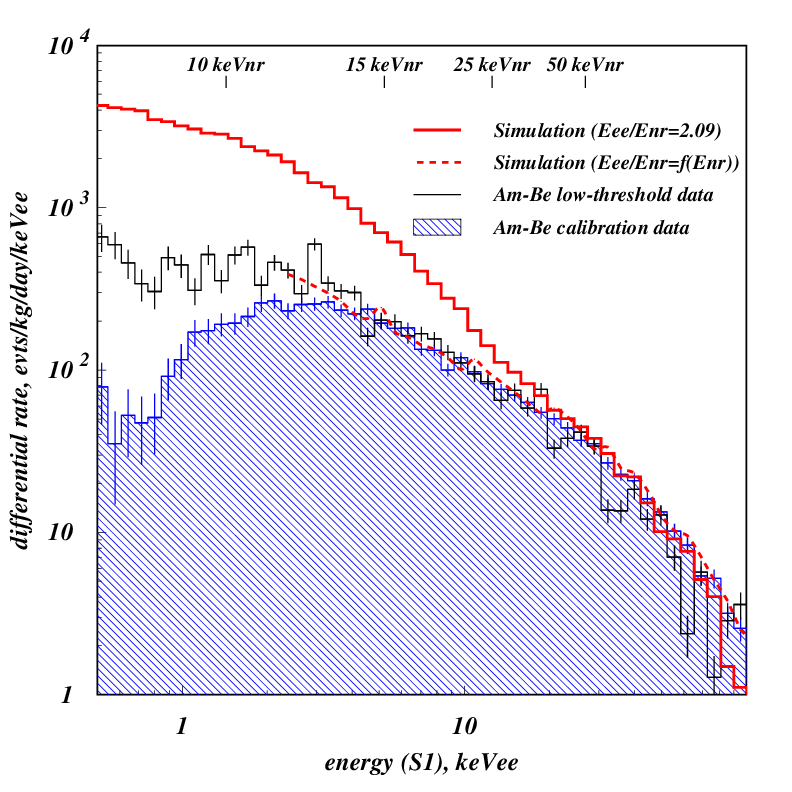}
\caption{\label{ambe_spec}Differential energy spectra for the AmBe
elastic recoil population in S1 electron-equivalent units
($^{57}$Co-calibrated S1). The main calibration data (shaded blue
histogram) and the lower threshold data-set described in the text
(black) are compared with the Monte Carlo simulation using a
constant conversion factor between nuclear recoil and electron
equivalent energies (solid red curve). The ratio between these
two curves is interpreted as an energy-dependent efficiency
factor and occurs in the low-energy region where thresholding effects are expected. The dashed red curve is the result of the nuclear recoil non-linearity
analysis described in the text, which results in the energy
conversion indicated by the markers at the top of the figure.}
\end{figure}

A comparison between  the differential spectrum seen during the nuclear-recoil
calibration, using AmBe, and a Monte Carlo simulation has been used to derive the energy scale conversion between keVee and keVnr. This relies on the integrity of the simulation using
GEANT4, which is  very well established in general for elastic scattering of
neutrons, and which has been further extensively validated as part of this work. Systematic effects related to the simulation of the experimental calibration were assessed. These included, amongst others: variations in neutron source spectrum and source location inside the shield; the effect of intervening and surrounding materials; simulation event selection; energy resolution smearing; coincident AmBe $\gamma$-rays; treatment of inelastic scattering in xenon. The Monte Carlo result at low recoil energies was very resilient to sensible variation of these parameters. A different Monte Carlo code \cite{FAUST} confirmed these results independently. Naturally, incorrect angular cross-sections for elastic scattering off xenon could be invoked to explain the low-energy result, since enhancing forward scattering would soften the recoil spectrum. However, dedicated simulations confirmed the correct implementation of the ENDF/B-VI evaluated data libraries \cite{ENDF} which underpin the GEANT4 low-energy neutron transport models. Both angular and energy-differential cross-sections were found to be in agreement with ENDF/B-VI and similar databases. An implementation in GEANT4 of the more recent ENDF/B-VII data for xenon by the XENON10 team~\cite{sore08}, aimed at exploring the causes of a similar effect observed by that experiment, found only minor differences in the recoil spectrum produced by a similar neutron source.  We have independently confirmed this conclusion.  The
comparison between simulation and experiment for ZEPLIN-III is shown in Figure~\ref{ambe_spec}. The energy scale
associated with the simulated data has been converted from keVnr to
keVee in Figure~\ref{ambe_spec} by simply dividing by 2.09, to allow
for the combination of the relative nuclear-recoil scintillation
efficiency to that of a 122\,keV $\gamma$-ray at zero electric
field, $L_{eff}$, and a suppression factor, $S$, which allows for
the field-dependent variation in the scintillation output.  These
are used in the following equation:
\begin{equation}
E_{nr}=\frac{S1}{L_y}\frac{S_e}{L_{eff}S_n}\,,
\end{equation}
where $S_e$ and $S_n$ are the suppression factors in the
scintillation output for 122\,keV $\gamma$-rays and nuclear recoils,
respectively, at the experiment operating fields. Note that in this
equation the ratio $S1/L_y$ defines the keVee unit. Above
$E_{nr}$$\sim$20~keV the available experimental data for $L_{eff}$
suggests it is constant at
$\sim$0.19~\cite{akimov02,aprile05,chepel06}.
  However a variation in $L_{eff}$
at low energy has been invoked to explain XENON10 neutron calibration data~\cite{sore08} and hence is allowed to vary in this work. 
\begin{table}
\caption{\label{efftab}Energy-independent efficiency factors and
thresholds due to hardware and software actions.  Efficiency figures
are constant over the WIMP recoil range.  Numbers following the
entries refer back to the list of software operations itemised in
Section~\ref{cuts}.  The total effective exposure is
127.8~kg$\cdot$days.}
\begin{tabular}{|l|c|l|}
\hline
{\bf Effect}&{\bf Efficiency}&{\bf Method}\\
\hline Deadtime&91.7\%&Measured \\
Hardware upper threshold&100\%&On-off compare\\
ZE3RA pulse finding (1) &96.0\%&Visual  inspection\\
&&Hand calculation\\
Event reconstruction (2,3) &91.9\%&Visual inspection\\
Selection cuts (5) &73.0\%&On-off compare\\
WIMP box acceptance&47.7\%&Calculation\\ \hline
{\bf Effect}&{\bf Threshold\footnote{All thresholds are quoted here in terms of the S1 signal in keVee for nuclear recoils.  The equivalent
nuclear recoil energy, keVnr, depends on the conversion between
keVee and keVnr. For the relationship shown in Section~\ref{ec}, 11
ionisation electrons corresponds to $<$~7keVnr}}&{\bf Method}\\
\hline Hardware (S2) trigger& S1$=0.5~$keVee
&Two data-sets\\
&&Visual inspection\\
&&Modeling\\
&&Pulser tests \\
Software S2 area &S1$<$1~keVee&Calculation\\
&&Scatter plots\\
Software S1 3-fold&S1$=1.7~$keVee&Calculation\\
&&Two data-set \\
&& analyses\\
 \hline
\hline
\end{tabular}
\end{table}

\begin{figure}
\includegraphics[width=3.5in,clip=]{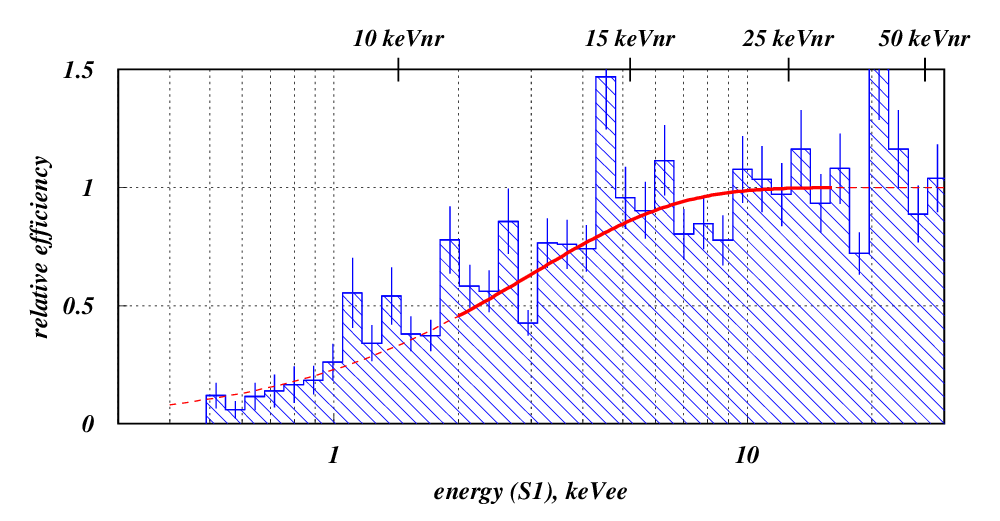}
\caption{\label{nlin}Energy-dependent part of the nuclear recoil
detection efficiency as deduced empirically by comparing the two experimental
AmBe spectra shown in Fig.~\ref{ambe_spec}. The `low-threshold' run
was taken with a lower hardware trigger threshold; in addition,
software quality cuts were relaxed, along with the S1 3-fold
requirement. A fit to the data is shown, with the WIMP
acceptance box indicated by the thicker portion of the line. The S2 hardware  trigger in the low-threshold AmBe run was half of that used in the science data-set and thus corresponded to S1\,=\,0.25\,keVee, and the S1 pulse finding algorithms only required a 2-fold detection above 1/3 p.e. giving a nominal software threshold of S1\,=\,1.1\,keVee.   Hence above S1\,=\,2\,keVee (the lower WIMP\ acceptance box boundary) the `low-threshold' data-set has near-unity efficiency.     }
\end{figure}
In general the conversion between an electron-equivalent
energy scale, in keVee, and a nuclear recoil energy scale, in keVnr,
is not necessarily linear and any non-linearity can be expressed
mathematically through energy dependency in $L_{e\!f\!f}$ and/or
$S_e/S_n$. Above $E_{nr}$$\sim$$20$~keV the available experimental
data for $L_{e\!f\!f}$  suggests it is constant at $\sim$0.19. At
lower energies the situation is much less clear~\cite{sore08}.  For
$S_n$ there are no data on the energy dependence but rather there is
a single value based on a measurement at 56~keVnr using a neutron
beam~\cite{aprile05}.  This gives $S_n = 0.90$ at our field and it
is commonly assumed to remain constant over the whole energy range
of WIMP nuclear recoils. If $L_{e\!f\!f}$ and/or $S_n$ are not
constant below $\sim$20~keVnr this will cause a non-linearity in the
nuclear recoil energy scale. 

\begin{figure}
\includegraphics[width=3.5in,clip=]{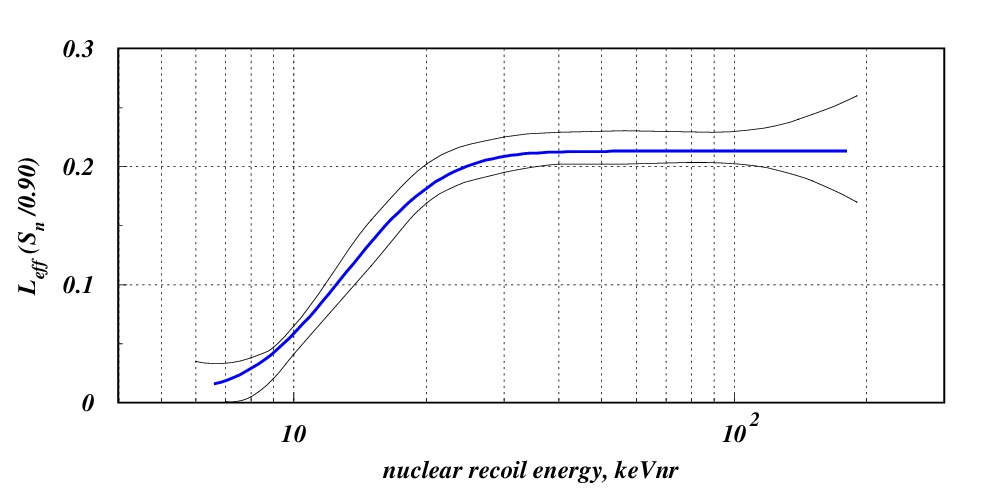}
\caption{\label{nlin2} The derived energy-dependent behaviour of
$L_{eff}\cdot S_n$. The thick curve shows the best fit to the data,
but other curves producing very similar goodness-of-fit indicators
are obtained within the envelope shown. The constraints become very
weak outside the energy range shown.}
\end{figure}

In the following it is assumed that
such non-linearities are responsible for the mismatch seen in
Figure~\ref{ambe_spec}.  The approach used is similar to that
applied to the XENON10 data~\cite{sore08}. Using a
maximum-likelihood technique we have derived a non-linearity
function which best matches the AmBe simulation to our neutron
calibration spectrum above $\sim$2~keVee. The outcome of this
process is shown as the dashed red curve in Figure~\ref{ambe_spec}.
Figure~\ref{nlin2} expresses the nonlinearity in terms of the
combined effect of $L_{e\!f\!f}$ and $S_n$, with the latter referenced to 0.90. In
Figures~\ref{ambe_spec}~and~\ref{nlin} the top horizontal axes show
the energy scale in keVnr to be compared with keVee on the bottom
scale.  The WIMP search box boundaries then translate to 10.7 and
30.2~keVnr.  One consequence of the required non-linearity is a
marked reduction in efficiency for nuclear recoil detection below
15~keVnr.

\subsubsection{Limit analysis}
The event box contains a large empty region with a small number of
events close to where a tail from the electron recoil distribution
is expected. However, although there is a good fit of a skew-Gaussian distribution to the electron-recoil band above the WIMP search box, there remains systematic uncertainty about an extrapolation of this being used as an accurate estimator of the number of expected background events in the box.  The fact that the best fit expectation exceeds the measured number of events might result in an artificially lower upper limit, as pointed out in \cite{cow98}.   This compromises any straightforward use of maximum-likelihood techniques and even the commonly-used Feldman-Cousins (FC)\ analysis \cite{cow98}.  Hence, a simpler, more transparent and conservative approach is adopted based on a minimum of three pieces of information about the data.
   
\begin{figure}
\includegraphics[width=3.5in,angle=0,clip=]{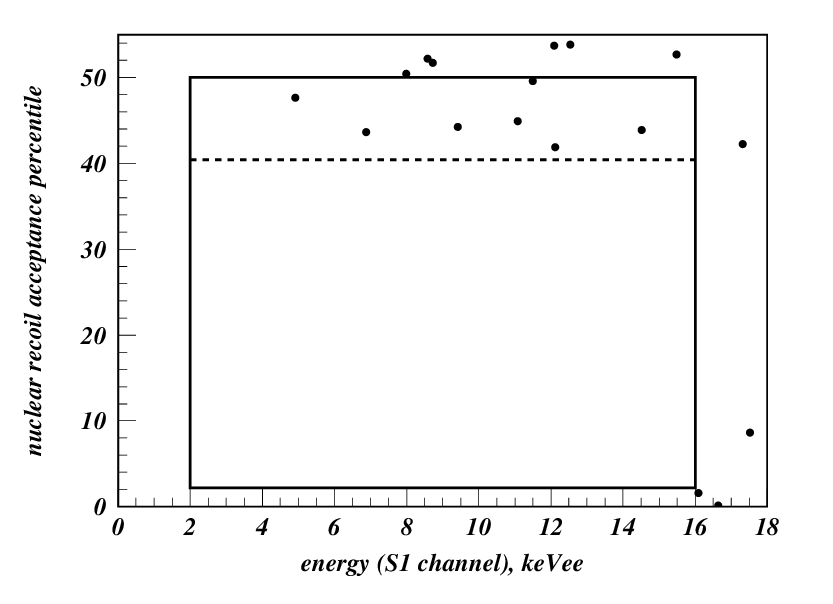}
\caption{\label{bins}The WIMP\ search box with the vertical axis remapped onto nuclear recoil percentiles.  This is done using the $S2/S1$ distribution from the AmBe calibration data.  The positions of the 7 events falling  within the box are shown as well as other events just outside the box. The horizontal dashed line separates the box into two regions with an area ratio of 1:4. }
\end{figure}

The first is the reasonable assumption that any expected electron-recoil background will fall in the top part of the WIMP search box.  Based on this assumption the box is divided into two regions which have significantly different probabilities of having electron-recoil background within them. This is done in Figure~\ref{bins} after transforming the WIMP\ search box so that the vertical axis has a linear scale in nuclear recoil acceptance percentiles as derived from the AmBe calibration data.   In this representation any WIMP\ nuclear recoil signal should populate the box uniformly, whereas the density of the electron recoil background is expected to decrease monotonically down from the top. A horizontal dashed line is shown which divides the WIMP search box into two regions such that the top area contains all the events.   In the following analysis the fractional area in the lower region is denoted by $f$.

The second is the observation that no WIMP\ event is seen in the lower region ($n_l=0$).

Finally,  it is possible that there may be up to 7 WIMP events in the upper region ($n_u\leqq 7$).

A classical 90\% one-sided upper limit for the WIMP expectation value in the whole box, $\mu$, is the value under which 10\% of repeated experiments would return zero  events in the lower box and up to 7 in the upper box.  This is expressed in terms of Poisson probabilities as 
\begin{eqnarray}
\lefteqn{P(n_l=0,n_u\leqslant7|\mu ) =} \nonumber \\
& & P(n_{l}=0|f\mu)*\sum_{i=0}^{7}
P(n_{u}=i|(1-f)\mu) = 0.1
\end{eqnarray}
Over the range of values of $f$ between 0.75 and 0.84 the calculated result is $\mu=2.30/f$. $f=0.84$ is the maximum area allowed which just excludes all of the events.

It turns out that, for the value of $\mu$ resulting from   this calculation, the second factor in equation (2) is very close to unity regardless of the area fraction, $f$.  This reflects the fact that the upper limit is driven almost entirely by the presence of the empty region and the value 2.30 is then recognised as the classical 90\% upper limit on zero. It is then reasonable to assume that the two-sided 90\% confidence interval for this particular data-set will also be driven by the empty box.  In this case  the upper limit to this interval will be at $\mu=2.44/f$, with 2.44 being the corresponding 2-sided FC\ upper limit on zero \cite{cow98}. Figure \ref{bins} shows a dividing line with $f=0.8$, which is adopted as a conservative boundary placement beyond which no background is likely. The 90\% confidence interval upper limit is then $\mu=3.05$.  With this extreme value of $\mu$  there is a 54\% probability that there are indeed no WIMP\ events in the upper region, a 33\% chance of there being 1 WIMP event and  a 13\% chance of $\geqq2$ WIMP events.  The fact that the most likely scenario is no WIMPs in the data-set even  with $\mu=3.05$ implies that   $\mu=0$ is included within the 90\% two-sided interval as the null event hypothesis becomes more and more likely as $\mu$ is reduced.

The upper limit of 3.05 events is used to derive the upper limit to the WIMP-nucleon spin-independent elastic scattering cross-section as a function of WIMP mass.  The signal energy distribution is obtained from the theoretical WIMP recoil spectrum~\cite{lew96}, derived using the standard  spherical
isothermal Galactic halo model ($\rho_{dm}$=0.3~GeVcm$^{-3}$,
$v_o$=220~km/s, $v_{esc}$=600~km/s and $v_{Earth}$=232~km/s),
detector response efficiencies and energy resolution.  The form
factor is taken from~\cite{helm56}.
  The expected distribution in $S2/S1$ is determined from the neutron calibration.

\begin{figure}
\includegraphics[width=3.5in,angle=0,clip=]{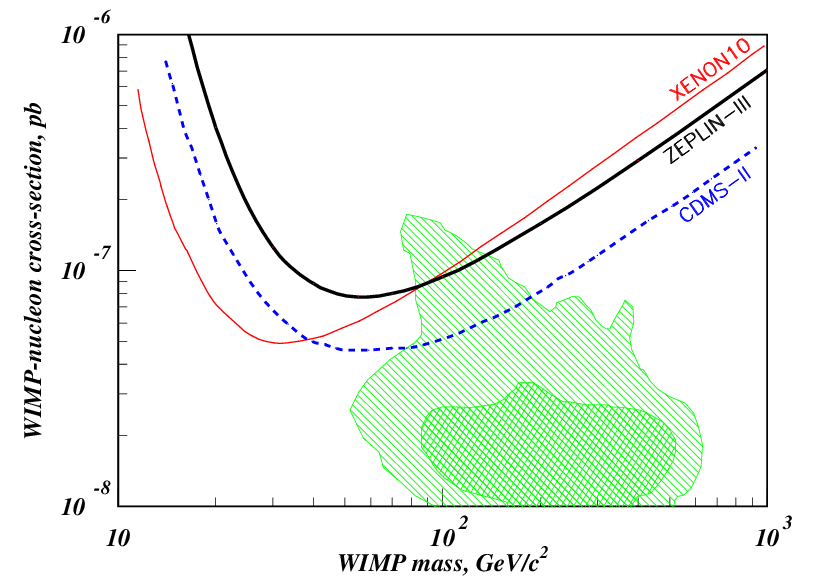}
\caption{\label{res}90\% confidence interval upper limit to the WIMP-nucleon
elastic scattering cross-section as derived from the first science
run of ZEPLIN-III for a spin-independent interaction. For
comparison, the experimental results from
XENON10~\cite{angle08,aprile08} and CDMS-II~\cite{akerib06} are also
shown.  Note that the XENON10 curve is a 1-sided limit,
corresponding approximately to an 85\% confidence 2-sided
limit~\cite{angle08}. CDMS-II and our result are both 90\% 2-sided
limits.  The hatched areas show 68\% and 95\% confidence regions for the neutralino-proton scattering cross-section with flat priors as calculated in Constrained MSSM \cite{trot08}. }
\end{figure}

 The final result for the 90\% confidence interval upper limit to the cross-section, shown in
Figure~\ref{res}, has a minimum  of
8.1$\times$10$^{-8}$~pb for a WIMP mass of 60~GeV/$c^2$. In the mass
range beyond 100~GeVc$^{-2}$ this result complements the XENON10
result and further constrains the favoured SUSY parameter
space~\cite{ros07} from xenon-based experiments.  Spin-dependent
limits are presented separately~\cite{leb09}.
\section{Conclusions}
An analysis of 847~kg$\cdot$days of data from the first science run
of ZEPLIN-III has resulted in a signal lower limit consistent with
zero, and an upper limit on the spin-independent WIMP-nucleon
elastic scattering cross-section of $8.1 \times 10^{-8}\,$pb, at
90\% confidence level.  In reaching this result it was necessary to
confront an unexpected mismatch between the nuclear recoil spectrum
shown in the AmBe calibration data and the Monte Carlo simulation. A
careful and thorough analysis of efficiency factors and threshold
effects (including the use of alternative data-sets with different
thresholds, systematic changes to software cuts and thresholds,
visual scanning and manual analysis of large samples of data and
modelling and direct verification of the performance of the DAQ) did
not resolve this mismatch.  As a more credible alternative
explanation the possibility of a non-linearity in the nuclear recoil
energy scale has been studied. Non-linearity as such is not
unexpected and, indeed it would be surprising if it did not exist at
low energy, and a similar approach has been used by others for
xenon~\cite{sore08}. Using this analysis it has been possible to
reconcile the data with a non-linearity setting in at the same
energy as in \cite{sore08} but with a more significant effect at lower energies. In itself this may not be surprising given
the very different operating conditions within ZEPLIN-III and
XENON10: the most obvious being that the electric field in the
liquid is 6 times stronger in the former. Indeed, there are other clear
differences in the performances of the two instruments. However, it
is clear that the physics underlying the low-energy performance is
poorly understood. This is true of both the response to electron
recoils~\cite{shutt07} and to nuclear recoils~\cite{sore08}. As a
point of reference, if the mismatch between the AmBe simulation and
the data were interpreted solely as an instrument efficiency, the
effect on the upper limit would not have been dramatic ($<$40\%
increase) as this approach has a better effective threshold for
nuclear recoils but a poorer efficiency.

The analysis presented is not blind as one of the analysis routines
was changed after opening of the full data-set as was the limit setting procedure.  In applying the limit analysis no use was made of any background estimates (neither electron-recoil or neutron scattering) and this was done deliberately to avoid underestimating the upper limit.  
\section*{Acknowledgements}
The UK groups acknowledge the support of the Science \& Technology
Facilities Council (STFC) for the ZEPLIN-III project and for
maintenance and operation of the underground Palmer laboratory which
is hosted by Cleveland Potash Ltd (CPL) at Boulby Mine near
Whitby, on the North-East coast of England. The project would not be
possible without the cooperation of the management and staff of CPL.
We also acknowledge support from a Joint International Project
award, held at ITEP and ICL, from the Russian Foundation of Basic
Research (08-02-91851 KO~a) and the Royal Society. We are indebted
to our colleagues at ITEP, D.Yu Akimov, V. Belov, A. Burenkov and A.
Kobyakin for their contributions. LIP-Coimbra acknowledges financial
support from Funda\c{c}\~{a}o para a Ci\^{e}ncia e Tecnologia (FCT)
through the project-grants POCI/FP/81928/2007 and
CERN/FP/83501/2008, the postdoctoral grant SFRH/BPD/27054/2006, as
well as the PhD grants SFRH/BD/12843/2003 and SFRH/BD/19036/2004.
The University of Edinburgh is a charitable body, registered in
Scotland, with the registration number SC005336.


\end{document}